# Observation of Dual Spin Reorientation Transitions in Polycrystalline CeCr$_x$Fe$_{1-x}$O$_3$ (x=0.33 and 0.67)

Stephen Tsui, Department of Physics, California State University San Marcos, 333 S. Twin Oaks Valley Road, San Marcos, CA 92096, USA.

Sara J. Callori, Department of Physics & Astronomy, California State University San Bernardino, 5500 University Parkway, San Bernardino, CA 92407, USA.

## Abstract

We investigate the magnetic behavior of polycrystalline CeCr$_x$Fe$_{1-x}$O$_3$ (x = 0, 0.33, 0.67, and 1) synthesized via solid state reaction. Rare earth orthoferrite and orthochromite materials are well known for exhibiting spin reorientation transitions. Cr$^{3+}$ doping in CeFeO$_3$ results in the unusual occurrence of two spin reorientation transitions, $\Gamma_4(G_x, A_y, F_z) \rightarrow \Gamma_1(A_x, G_y, C_z)$ near 230 K and $\Gamma_4(G_x, A_y, F_z) \rightarrow \Gamma_2(F_x, C_y, G_z)$ near 100 K. In addition, two Néel transitions are identified. The results indicate that CeCr$_x$Fe$_{1-x}$O$_3$ offers a rich collection of magnetic behaviors with application potential for spintronic devices.

## Introduction

There exists an extensive body of literature on the magnetic behavior of the rare earth orthoferrites and orthochromites RMO$_3$, where R= rare earth element and M=Fe or Cr, respectively. These materials crystallize in a distorted orthorhombic structure and exhibit a canted *G*-type antiferromagnetic ground state [1]. Interactions of the transition metal and rare earth ions in these systems lead to interesting behaviors including magneto-optical effects [2,3], magnetization reversal [4,5], multiferroicity [6-8], photocatalysis [9, 10], and most notably spin reorientation (SR) [1, 11-12]. The magnetic behaviors of these RMO$_3$ systems originate from the competing exchange interactions between the crystal ions: R$^{3+}$-R$^{3+}$; R$^{3+}$-M$^{3+}$; and M$^{3+}$-M$^{3+}$ [11, 12].

These magnetic interactions lead to three allowed spin configurations in the rare earth orthoferrites and orthochromites as described by components of the ferromagnetic vector *F* and antiferromagnetic vectors *A*, *C*, and *G*: $\Gamma_1(A_x, G_y, C_z)$, $\Gamma_2(F_x, C_y, G_z)$, and $\Gamma_4(G_x, A_y, F_z)$. The notation was established by Wollan and Koehler [13] and Bertaut [14]. Per Wollan and Koehler [13, 15], *A*-type antiferromagnetism arises from a cubic unit cell where the ions in the bottom face *a*-*b* plane possess the same spin orientation as those in the bottom plane but opposite spin orientations from the ions in the top face *a*-*b* plane, creating a structure of alternating layers. *C*-type antiferromagnetism is described by a cubic unit cell whereby each *a*-*c* plane possesses two ions of the same spin orientation in the *c*-direction on one edge and two ions of opposing spin orientation on the next edge, essentially creating "strands" of alternating opposing spins. Lastly, *G*-type antiferromagnetism results from opposite spins for each nearest neighbor along the cubic directions which results in similar spins positioned diagonally across each cube face. In summary, $\Gamma_4(G_x, A_y, F_z)$ denotes a small net magnetic moment along the *c*-axis due to canting of the $G_x$ spins. The antiferromagnetic $A_y$ spin arrangement also occurs due to $G_x$ canting.

$\Gamma_2(F_x, C_y, G_z)$ indicates a small net magnetic moment along the $a$-axis with the $C_y$ representing another canted spin arrangement. Lastly, $\Gamma_1(A_x, G_y, C_z)$ describes a compensated spin configuration with a collinear antiferromagnetic structure where there is no net magnetic moment [1, 11]. The SR phenomenon, whereby the easy axis of magnetization changes as a result of transitions between the above spin configurations due to the $R^{3+}$-$M^{3+}$ interaction, has long garnered strong interest in these materials.

In particular, $CeFeO_3$ has been studied for its relatively high SR transition temperature ~230-240 K [16, 17]. The synthesis of $CeFeO_3$ is mentioned in the literature as early as 1956 [18], and it is noted that the easy oxidation of $Ce^{3+}$ to $Ce^{4+}$ makes the synthesis a challenge [16, 19-20]. For $CeFeO_3$, the spin configuration is $\Gamma_4$ at room temperature as determined by the $Fe^{3+}$ magnetic sublattice [15]. The Néel transition temperature for $Fe^{3+}$ ions typically ranges from $T_N = 620$-740 K [17]. The $Ce^{3+}$-$Fe^{3+}$ interaction causes the SR transition near 240 K as the system undergoes a first order $\Gamma_4 \rightarrow \Gamma_1$ transition [15-17]. The $\Gamma_1$ state is dictated by the $Ce^{3+}$ magnetic sublattice [15].

In contrast to its cousin compound $CeFeO_3$, $CeCrO_3$ is perhaps best regarded for its magnetization reversal behavior due to antiferromagnetic coupling between the $Ce^{3+}$ and $Cr^{3+}$ magnetic moments [5, 21]. As reported by Cao et al. [5], below the Néel transition temperature $T_N = 260$ K, the $CeCrO_3$ spin configuration is $\Gamma_4$ and is dominated by the $Cr^{3+}$ magnetic sublattice. Under field-cooled cooling (FCC) magnetic measurement conditions, the magnetization increases as the sample is cooled to a maximum near 233 K before decreasing. This decrease in the magnetization reaches a compensation point at 133 K. The spin flipping arises from the polarization of the $Ce^{3+}$ ions antiparallel to the combination of the internal field of the $Cr^{3+}$ ions and the applied external magnetic field [22]. Further cooling leads to a second order spin reorientation of the $Cr^{3+}$ moments, resulting in a $\Gamma_2$ configuration near 16 K. Under zero field-cooled (ZFC) magnetic measurement conditions, no such spin flip occurs [5, 22].

Several studies have investigated the effects of rare earth substitution such as $Cr^{3+}$ for the $Fe^{3+}$ ions on the SR transition for various compounds including $HoFeO_3$ [23], $SmFeO_3$ [24], and $PrFeO_3$ [25]. Recent literature describes the magnetic behavior of Fe-doped $CeCrO_3$ nanoparticles synthesized using a combustion method [22, 26]. In a similar vein, the purpose of this work is to investigate the magnetic behavior of stoichiometrically Cr-doped $CeFeO_3$ polycrystalline bulk samples, which to the authors' knowledge is not reported elsewhere in the literature.

## Methods

Polycrystalline $CeCr_xFe_{1-x}O_3$ samples were synthesized via the following stoichiometric solid state reactions:

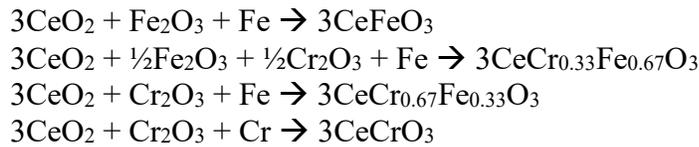

$3CeO_2 + Fe_2O_3 + Fe \rightarrow 3CeFeO_3$
$3CeO_2 + \frac{1}{2}Fe_2O_3 + \frac{1}{2}Cr_2O_3 + Fe \rightarrow 3CeCr_{0.33}Fe_{0.67}O_3$
$3CeO_2 + Cr_2O_3 + Fe \rightarrow 3CeCr_{0.67}Fe_{0.33}O_3$
$3CeO_2 + Cr_2O_3 + Cr \rightarrow 3CeCrO_3$

The materials used were $CeO_2$ (99.9% Thermo Scientific), $FeO_2$ (99.995% Thermo Scientific), Fe (99% Thermo Scientific), $Cr_2O_3$ (99.97% Thermo Scientific), and Cr (99% Thermo Scientific). The precursor powders were ground using a mortar and pestle before being pressed into pellets under a pressure of approximately 82 kPa in a hydraulic press. The pellets were placed into a Lindbergh Blue quartz tube furnace and then kept under an approximately 10 μTorr vacuum via an Alcatel MDP 5011 molecular drag pump. The samples were heated to 1100 Celsius over a 7 hour temperature ramp, sintered for 99.5 hours, and then cooled to room temperature over 7 hours.

Sample crystal structure was determined via X-ray diffraction using a Rigaku SmartLab diffractometer using Copper $K_\alpha$ wavelength. Magnetic characterization of the samples was performed using a Quantum Design VersaLab Physical Properties Measurement System with Vibrating Sample Magnetometer (VSM) option. Zero-field cooled (ZFC) and field-cooled cooling (FCC) magnetic measurements were performed to investigate the temperature-dependent magnetization M(T) behavior from 50 K-300 K. The isothermal applied magnetic field-dependent magnetization M(H) behavior was measured at 50 K and 300 K.

**Results and Discussion**

Samples of $CeFeO_3$ and $CeCrO_3$ were first measured to establish structural information for the undoped compounds. The $CeFeO_3$ and $CeCrO_3$ X-ray data were well fit with the known orthorhombic *Pbnm* structure (Figure 1). It should be noted that *Pbnm*, rather than the similar *Pnma* structure, is typically used to describe rare earth orthoferrite systems because of its early adoption by Bertaut [15]. $CeCr_xFe_{1-x}O_3$ samples doped with both x =0.33 and x = 0.67 were compared to the two end members of the series to determine to what extent the Cr was incorporated into the orthorhombic $CeFeO_3$ structure rather than forming other distinct phases. X-ray diffraction data is shown for several samples in Figure 2. The majority of the peaks of the doped samples are indicative of a *Pbnm* structure, indicating that the Cr was integrated into the $CeFeO_3$ structure. There are several peaks that are attributed to $CeO_2$, one of the precursor materials. This is a known issue with the synthesis of $CeFeO_3$ [16, 19-20] and $CeCrO_3$ [26]. The relative peak intensities indicate that the final powder samples are approximately 88% $CeFeO_3$, 96% $CeCr_{0.33}Fe_{0.67}O_3$, and 80.8% $CeCr_{0.67}Fe_{0.33}O_3$. We note that the synthesis of $CeCrO_3$ was more challenging, with a single phase yield of 54%. Fortunately, the magnetic data to be later discussed is consistent with what has been reported in the literature, and because bulk $CeO_2$ is typically paramagnetic and possesses little magnetic moment, there is little contribution to the measured magnetic signal [27].

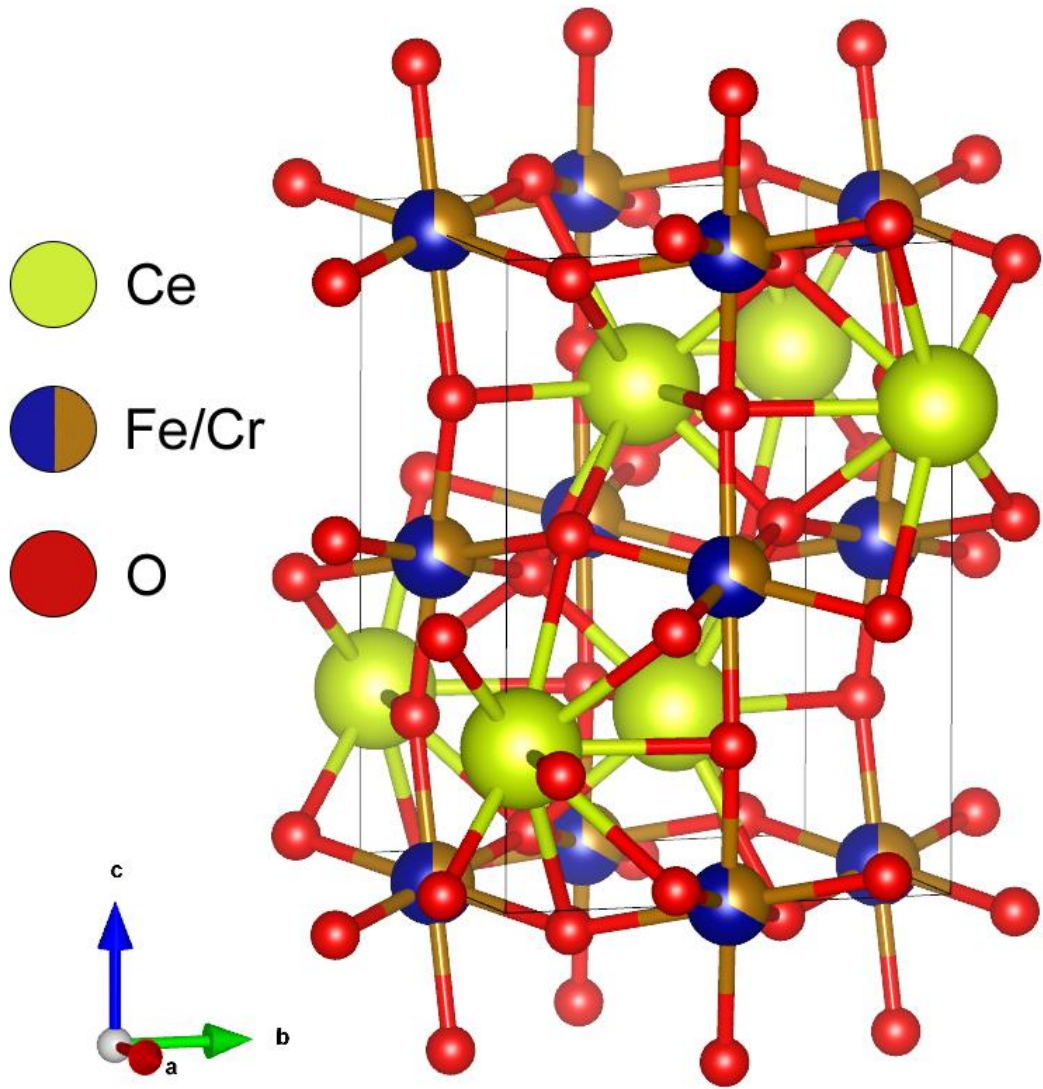

Figure 1. The *Pbnm* CeMO$_3$ structure, where M = Fe or Cr. The cerium ions are denoted in green, the M ions in blue/brown, and the oxygen ions in red.

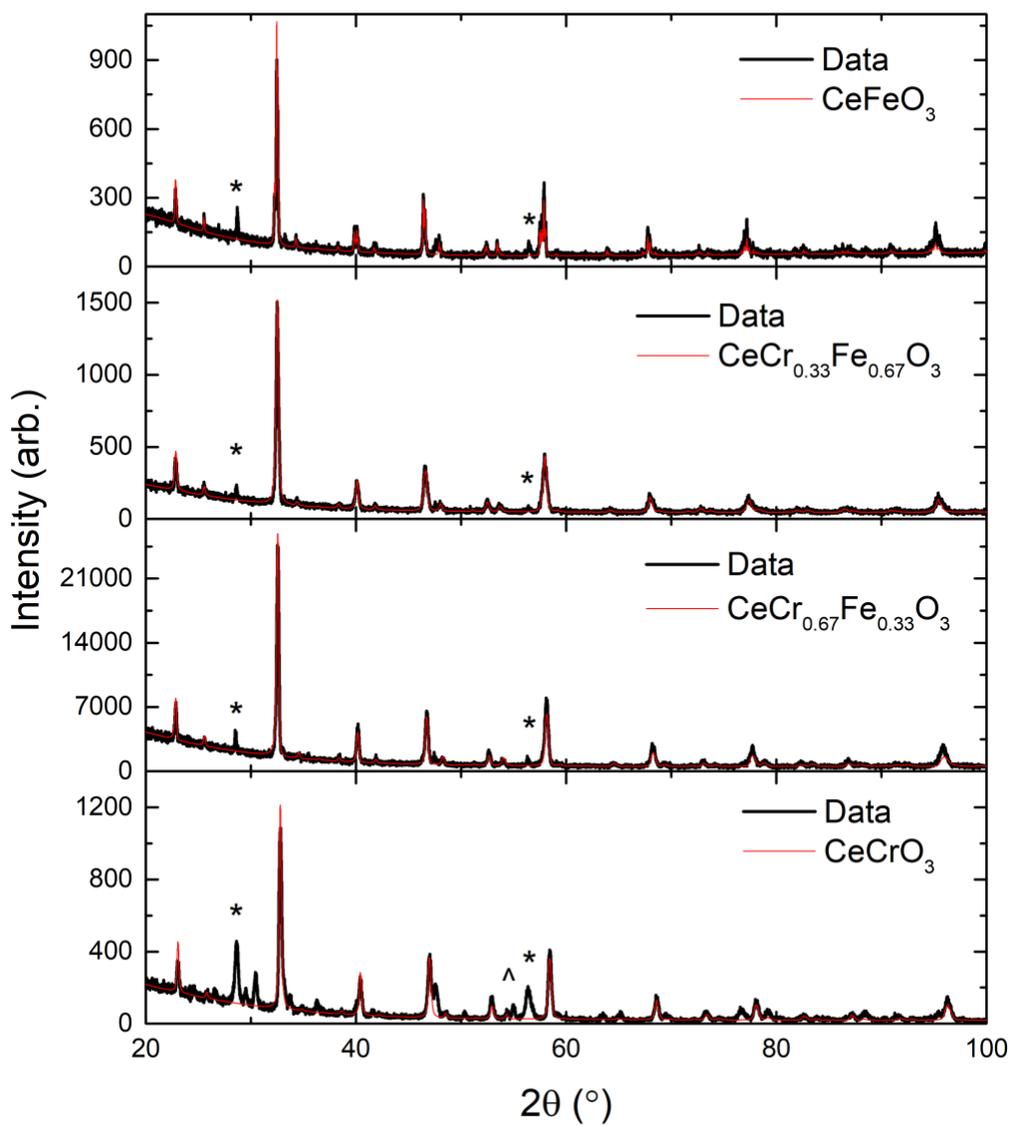

Figure 2. X-ray diffraction patterns showing experimental data (black lines) and fits (red lines) for: $CeFeO_3$, $CeCr_{0.33}Fe_{0.67}O_3$, $CeCr_{0.67}Fe_{0.33}O_3$, and $CeCrO_3$. Asterisks indicate peaks corresponding to unreacted $CeO_2$.

The ZFC and FCC magnetization vs. temperature M(T) curves measured at 1000 Oe for all samples are plotted in Figure 3. The CeFeO$_3$ sample's behavior is in good agreement with the literature [16, 17], prominently demonstrating the first order $\Gamma_4 \rightarrow \Gamma_1$ SR transition onset between $T_{SR1} = 232$ K and 235 K. The data also demonstrate a small drop in magnetization near $T_{SR2} = 132$ K, which we believe confirms Hou et al.'s [16] observation of an additional second order $\Gamma_4 \rightarrow \Gamma_2$ transition. In their work, they suggest that the presence of additional oxygen during the synthesis process leads to an increase in the ratio of $Ce^{4+}$ vs. $Ce^{3+}$ ions in certain regions of the sample, which in turns weakens the $Ce^{3+}$ - $Fe^{3+}$ interaction and permits $Fe^{2+}$ - $Fe^{3+}$ interactions to become more dominant. As a result, there is a coexistence of some $\Gamma_4$ and $\Gamma_1$ states below the 232 K $T_{SR1}$ transition. The availability of these additional $\Gamma_4$ states permits the second spin reorientation to $\Gamma_2$ at lower temperature $T_{SR2}$.

Similarly, the CeCrO$_3$ sample's behavior (Fig. 3) is also in good agreement with the literature [5, 21-22]. A canted antiferromagnetic transition occurs at the Néel temperature $T_N = 260$ K due to ordering of the $Cr^{3+}$ ions. Under FCC conditions, the magnetization increases to a peak near 221 K followed by a magnetization decrease until reaching a compensation point $T_{comp} = 100$ K at which the magnetization crosses over to become negative, due to the polarization of the $Ce^{3+}$ moments. Because of the temperature limit of our instrument, we were unable to examine the reported $\Gamma_4 \rightarrow \Gamma_2$ transition near 16 K.

The CeCr$_{0.33}$Fe$_{0.67}$O$_3$ sample (Fig. 3) exhibits a rich set of M(T) behaviors reminiscent of combined behaviors of CeFeO$_3$ and CeCrO$_3$. Cooling from room temperature, the magnetization begins to drop with an onset near 260 K, which coincides with $T_N$ in undoped CeCrO$_3$. Further cooling leads to a change in the slope of the magnetization drop near 230 K, which corresponds with $T_{SR1}$ in undoped CeFeO$_3$. Below $T_{SR1}$, our data is consistent with Yadav et al.'s nanoparticle results [22, 26]. They describe CeCr$_{1-x}$Fe$_x$O$_3$ as demonstrating two Néel transitions, one involving $Cr^{3+}$-$Fe^{3+}$ interaction at higher temperature and one involving $Cr^{3+}$-$Cr^{3+}$ at lower temperature. Utilizing their description, we observe a Néel transition $T_{N2} = 200$ K. Upon further cooling, the magnetization rises until reaching a broad peak near 100 K before further decreasing. Yadav et al. report a sharper peak at similar temperature range and attribute this behavior to the enhancement of the 16 K transition temperature for the $\Gamma_4 \rightarrow \Gamma_2$ SR transition that nominally occurs in undoped CeCrO$_3$ due to the $Ce^{3+}$-$Cr^{3+}$ interaction [5].

The CeCr$_{0.67}$Fe$_{0.33}$O$_3$ sample (Fig. 3, inset) possesses a significant bifurcation of the ZFC and FCC data at the Néel transition at $T_{N1} = 260$ K. For the moment, we focus on the ZFC data. As in the x = 0.33 sample, the $\Gamma_4 \rightarrow \Gamma_1$ SR transition occurs at $T_{SR1} = 230$ K. The data suggest that the first order $\Gamma_4 \rightarrow \Gamma_1$ transition temperature in CeFeO$_3$ is robust to the Cr-doping, which is not the case in $\Gamma_4 \rightarrow \Gamma_2$ systems such as HoFeO$_3$ where the presence of $Cr^{3+}$ interferes with the $Fe^{3+}$-$Fe^{3+}$ interaction, thereby increasing the SR transition temperature [23]. Upon further cooling, the data reveals $T_{N2} = 210$ K and the $\Gamma_4 \rightarrow \Gamma_2$ SR transition at $T_{SR2} = 90$ K in the ZFC data. The decrease in the $T_{SR2}$ temperature with increasing Cr content is consistent with reported observations in CeCr$_{1-x}$Fe$_x$O$_3$ nanoparticles [26]. However, below $T_{N2}$, the FCC data resembles the behavior of CeCrO$_3$ with a significant increase in the magnetization to a plateau near 60 K. Following the description of Shukla et al. [28], the $Ce^{3+}$ and $Cr^{3+}$ moments are uncoupled under ZFC conditions and individually contribute to the overall magnetization. Under FCC conditions, however, the canted antiferromagnetic ordering of the $Cr^{3+}$ ions leads to an increasing magnetization, which overcomes any indication of $T_{SR2}$. The rising strength of the $Cr^{3+}$ magnetization imposes a local magnetic field on the $Ce^{3+}$ ions. At a certain temperature, the $Ce^{3+}$ moments couple antiferromagnetically with the $Cr^{3+}$ moments and begin to dominate the magnetization, resulting in a local peak in the FCC data followed by a continued decrease in the net magnetization upon further cooling. Unfortunately, we cannot fully investigate this behavior below 50 K due to the limit of our instrumentation.

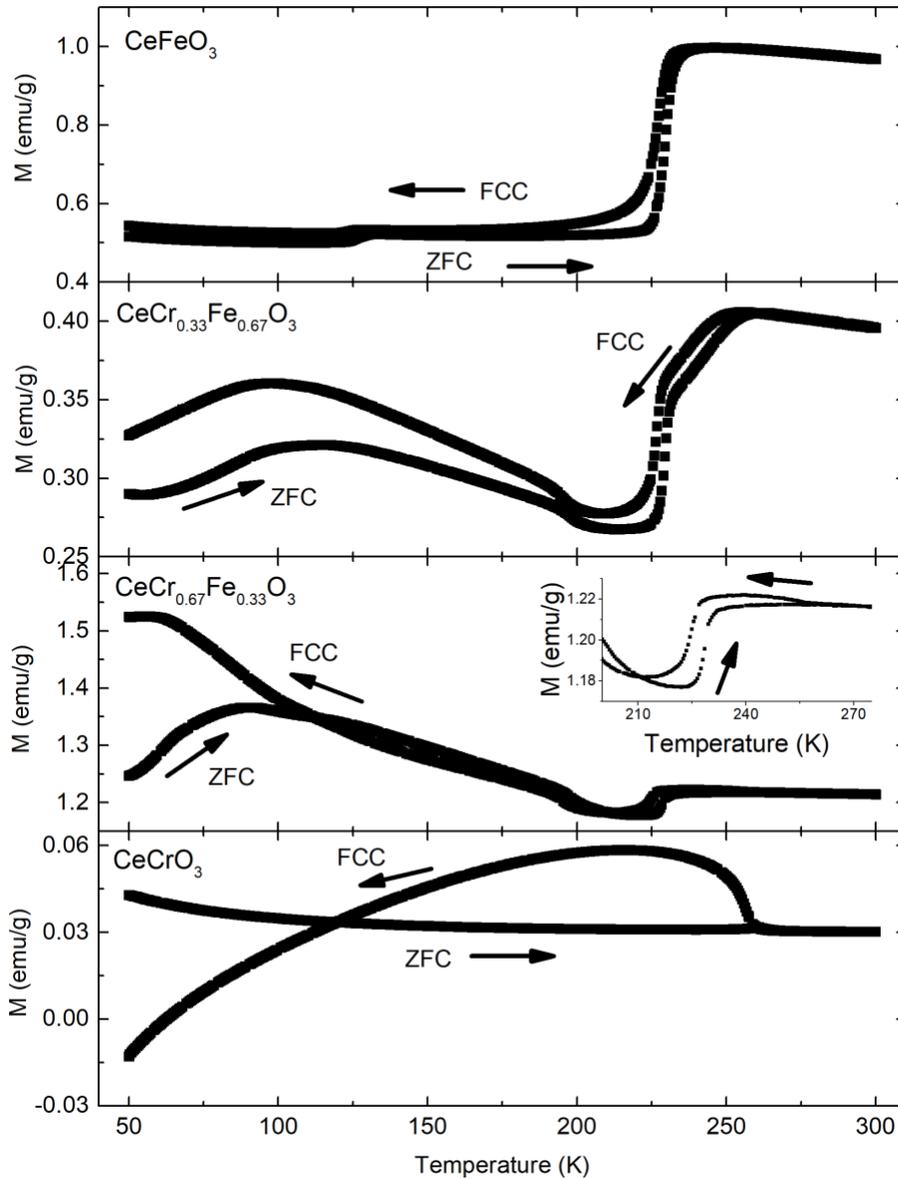

Figure 3. Magnetization vs. temperature for polycrystalline samples of: $CeFeO_3$, $CeCr_{0.33}Fe_{0.67}O_3$, $CeCr_{0.67}Fe_{0.33}O_3$, and $CeCrO_3$. All samples were measured using an applied magnetic field of 1000 Oe.

In Figure 4, ZFC and FCC M(T) data for the doped samples are taken at applied magnetic fields of 1000 Oe, 1 T, and 3 T. For the x = 0.33 sample, the $\Gamma_4 \rightarrow \Gamma_1$ SR transition is prominent at 1 T, as are the features at $T_{N1}$ and $T_{N2}$. The lower temperature peak attributed to $T_{SR2}$ is not observable under such strong applied magnetic field. Similarly, the x = 0.67 sample at 1 T does not display the $T_{SR2}$ = 90 K peak but does show the $T_{SR1}$ transition. Overall, at that high applied field strength, the ZFC and FCC curves behave more similarly. At 3 T, the most prominent feature for both samples is the Néel transition near $T_{N2}$ = 200 K, which demonstrates the strong contribution of the canted antiferromagnetic ordering of the $Cr^{3+}$ ions.

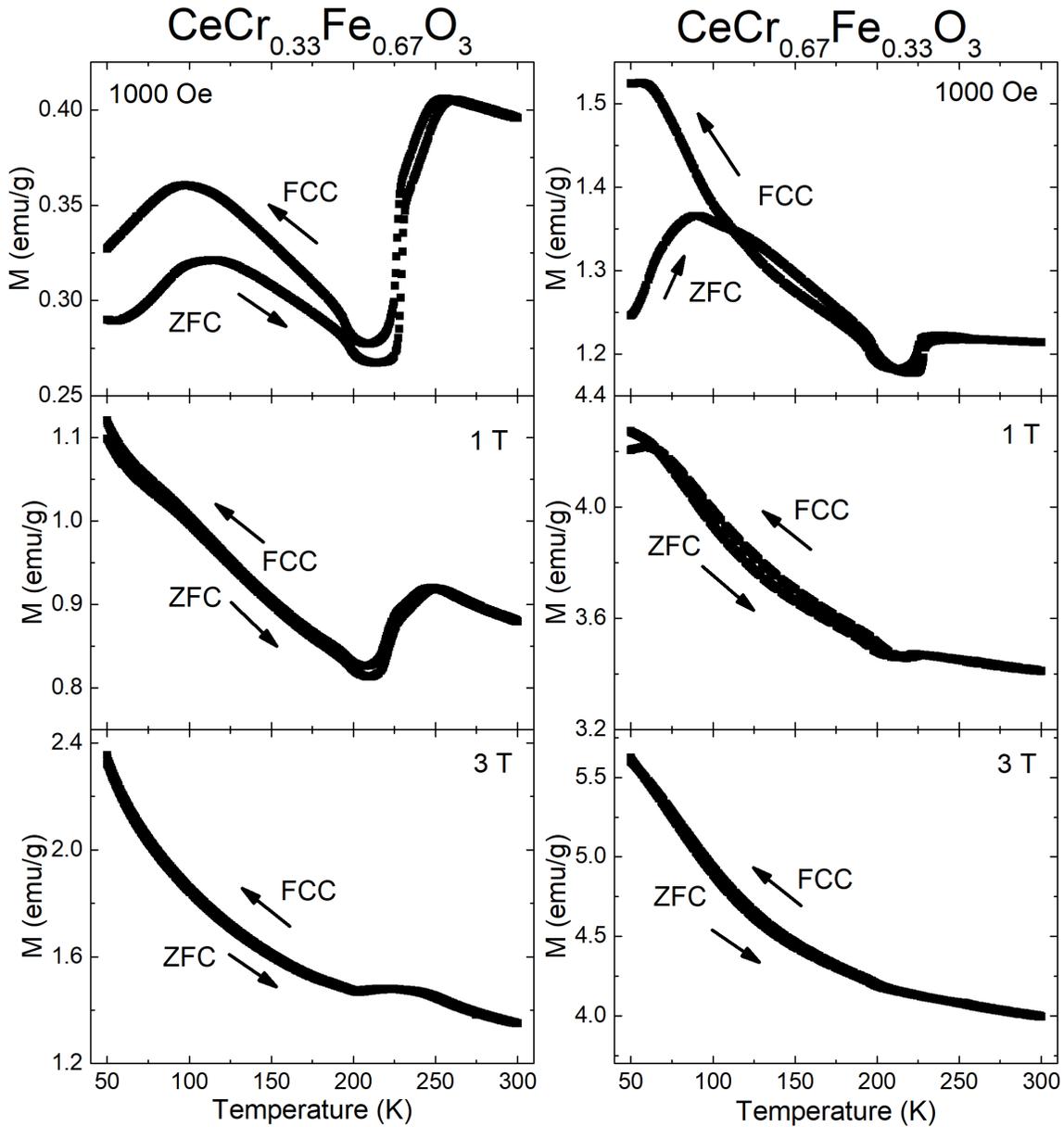

Figure 4. Magnetization vs. temperature M(T) at different applied magnetic field strengths for polycrystalline samples of CeCr$_{0.33}$Fe$_{0.67}$O$_3$ (left) and CeCr$_{0.67}$Fe$_{0.33}$O$_3$ (right).

It should be noted that there are no reports in the literature of $CeCr_xFe_{1-x}O_3$ ($x \leq 0.5$) nanoparticles exhibiting any $\Gamma_4 \rightarrow \Gamma_1$ SR transition. Nevertheless, the coexistence of two SR transitions in a single system is not unreasonable, given that multiple SR transitions have been reported in, for example, doped $DyFeO_3$ [29, 30] and that $HoFeO_3$ possesses additional metastable magnetic configurations as part of its spin reorientation transition [31]. We suggest qualitatively that there are two possible mechanisms that describe our observation of two spin reorientations in our doped samples.

The first possible mechanism simply assumes that the $\Gamma_4 \rightarrow \Gamma_1$ SR transition due to the $Ce^{3+}$-$Fe^{3+}$ interaction is incomplete, such that there is a coexistence of some $\Gamma_4$ and $\Gamma_1$ states below $T_{SR1}$. This is not dissimilar from Hou's attributed mechanism for two-fold SR transitions in $CeFeO_3$ [16]. The additional $\Gamma_4$ regions presumably contain $Cr^{3+}$ ions which can interact with $Ce^{3+}$ ions at lower temperatures, resulting in the $\Gamma_4 \rightarrow \Gamma_2$ SR transition at $T_{SR2}$.

The second possible mechanism assumes that the $\Gamma_4 \rightarrow \Gamma_1$ SR transition is complete and that the $\Gamma_1$ configuration persists below $T_{SR1}$ until the $Cr^{3+}$ moments order at $T_{N2}$. The strength of this canted antiferromagnetic ordering overcomes the $\Gamma_1$ configuration and puts the sample in $\Gamma_4$ configuration, which persists at lower temperature until the $Ce^{3+}$-$Cr^{3+}$ interaction begins the $\Gamma_4 \rightarrow \Gamma_2$ SR transition at $T_{SR2}$.

Figure 5 depicts the magnetization vs. applied magnetic field M(H) data for all samples under ZFC conditions at 50 K and 300 K. The $CeFeO_3$ sample displays a small hysteretic loop at 300 K and an unsaturated S-shaped curve at 50 K, as expected from the M(T) data and consistent with the literature [16, 17]. The behavior is similar for $CeCr_{0.33}Fe_{0.67}O_3$. For $CeCr_{0.67}Fe_{0.33}O_3$, the data demonstrate unsaturated S-shaped curves with a small hysteretic loop at 50 K. In the case of $CeCrO_3$, the data shows slight S-shaped curves without any hysteretic loops or indication of magnetic saturation, also in agreement with literature [21].

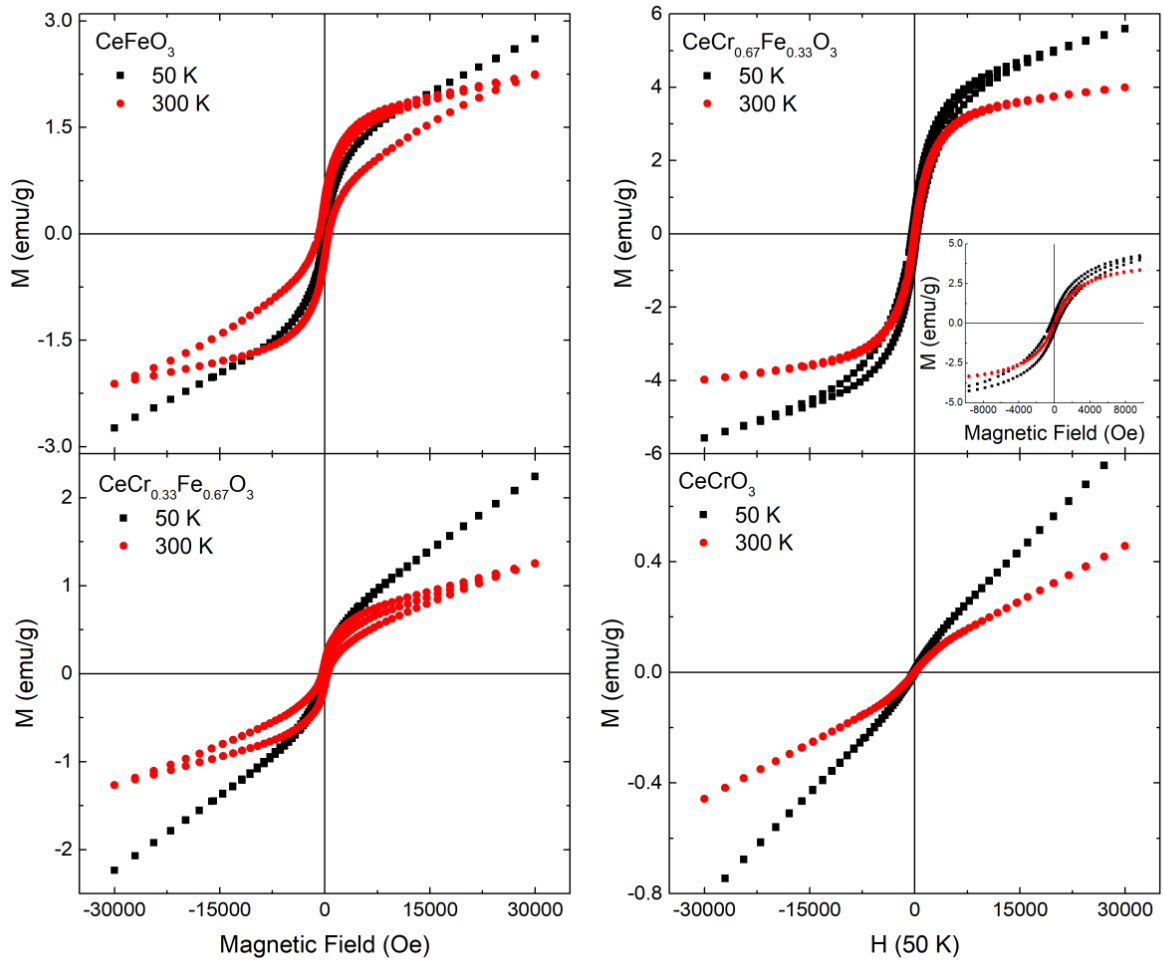

Figure 5. Magnetization vs. applied magnetic field M(H) for polycrystalline samples of $CeFeO_3$, $CeCr_{0.33}Fe_{0.67}O_3$, $CeCr_{0.67}Fe_{0.33}O_3$, and $CeCrO_3$.

## Conclusion

We have synthesized bulk polycrystalline $CeCr_xFe_{1-x}O_3$ (x = 0, 0.33, 0.67, and 1) with an orthorhombic *Pbnm* structure. $CeFeO_3$ possesses a spin reorientation transition $\Gamma_4(G_x, A_y, F_z) \rightarrow \Gamma_1(A_x, G_y, C_z)$ near 230 K, whereas $CeCrO_3$ is reported to undergo a $\Gamma_4(G_x, A_y, F_z) \rightarrow \Gamma_2(F_x, C_y, G_z)$ spin reorientation transition near 16 K. The introduction of Cr doping into $CeFeO_3$ results in the appearance of both the $\Gamma_4 \rightarrow \Gamma_1$ transition near 230 K and the $\Gamma_4 \rightarrow \Gamma_2$ transition associated with $CeCrO_3$ but at a greatly enhanced temperature near 100 K. $CeCr_{0.67}Fe_{0.33}O_3$ reveals bifurcation of the M(T) data between the ZFC and FCC conditions due to the strength of the antiferromagnetic ordering of the $Cr^{3+}$ ions in response to applied magnetic field. Observing two spin reorientation mechanisms in this system is remarkable, especially given that the literature to date only reports the existence of the single $\Gamma_4 \rightarrow \Gamma_2$ transition in $CeCr_xFe_{1-x}O_3$ nanoparticles.

## Acknowledgements

The work performed at California State University San Marcos was supported by funding from Quantum Design, Inc., through the Discovery Teaching Labs Initiative. The work performed at California State University San Bernardino was supported by funding from the NSF Centers for Excellence in Science and Technology (CREST) program (Grant #1914777) and the NSF Major Research Instrumentation (MRI) program (Grant #1920356). The authors would like to thank R. Dumas, M. Gooch, J. Perez, and J. Pham for their constructive feedback on this study.